# Active graphene plasmonics with a drift-current bias

Tiago A. Morgado[1], Mário G. Silveirinha[1,2*]

[1]*Instituto de Telecomunicações and Department of Electrical Engineering, University of Coimbra, 3030-290 Coimbra, Portugal*

[2]*University of Lisbon, Instituto Superior Técnico, Avenida Rovisco Pais, 1, 1049-001 Lisboa, Portugal*

*E-mail:* tiago.morgado@co.it.pt, mario.silveirinha@co.it.pt

**Abstract**

We theoretically demonstrate that a system formed by a drift-current biased graphene sheet on a silicon carbide substrate enables loss compensation and plasmon amplification. The active response of the graphene sheet is rooted in the optical pumping of the graphene plasmons with the gain provided by the streaming current carriers. The proposed system behaves as an optical amplifier for the plasmons co-propagating with the drifting electrons and as a strong attenuator for the counter-propagating plasmons. Furthermore, we show that the feedback obtained by connecting the input and output of the system, e.g., as a ring-shaped graphene – silicon carbide nanoresonator, combined with the optical gain provided by the drifting electrons, may lead to spasing.

---

[*] To whom correspondence should be addressed: E-mail: mario.silveirinha@co.it.pt

The unprecedented field enhancement and subwavelength confinement provided by surface plasmon polaritons (SPPs) [1] – charge density waves supported by metal-type surfaces – have pushed the field of plasmonics [2-4] to the frontline of scientific research. The unique features of the SPPs opened the door to a plethora of new phenomena and important applications, such as in nanophotonic circuitry [5-6], photonic metamaterials [7], solar energy harvesting [8], superlensing [9-10], chemical and medical sensing [11-13], and photothermal cancer therapy [14-15].

With the isolation of graphene [16] and the discovery of its remarkable electronic and optical properties [17-18], the field of plasmonics experienced a new boost [19-24]. Much of the interest in graphene plasmonics comes from the fact that the optical properties of this one-atom thick material are highly tunable by means of chemical doping or electrostatic gating, offering a unique opportunity to dynamically manipulate the SPPs properties.

Unfortunately, the high absorption (or ohmic) losses that intrinsically characterize plasmonic materials, such as metals and semiconductors, caused by different scattering mechanisms (e.g., electron-phonon and electron-electron scattering) and by Landau damping [25-26], impose harsh limitations in many nanophotonic applications. For instance, the ohmic losses in silver may limit the SPP propagation length to about $20\,\text{nm}$ at near-UV frequencies where the field confinement is strongest [27]. The propagation length increases to values up to $20\,\mu\text{m}$ for visible frequencies but at the expense of poor wave localization [1, 27]. The plasmonic dissipation in graphene is also quite significant [28-31], restricting the SPP propagation length to $1\,\mu\text{m}$ at mid-infrared frequencies and room temperature, and to about $10\,\mu\text{m}$ at cryogenic temperatures [32].

Even though the development of new plasmonic materials [33-35] may help mitigate the detrimental effects of ohmic losses, the ultimate limits imposed by plasmonic absorption (e.g., in the SPP propagation length or even in the resolution of superlenses) can be only surpassed by introducing optical gain into the systems. In this context, several theoretical and experimental studies on plasmonic loss compensation and SPP amplification have been reported [27, 36-54]. In particular, the amplification of long-range SPPs was experimentally demonstrated in systems formed by gold nanofilms combined with optically pumped gain media such as dye solutions [47] and fluorescent polymers [48]. Moreover, merging the SPP amplification with some feedback mechanism may lead to the spontaneous generation of SPPs, an effect known as spasing or plasmonic lasing [55-65].

In this work, we theoretically predict the full compensation of plasmonic loss and the amplification of SPPs in a nanostructure formed by a drift-current biased graphene sheet deposited on a silicon carbide (SiC) substrate (Fig. 1). The graphene-SiC plasmons gain energy from the electrons drifting on the graphene sheet, a process known as "negative Landau damping" [68]. It is shown that the considered graphene-SiC nanostructure acts as an amplifier for the SPPs co-propagating with the drifting electrons and as a very effective attenuator for the counter-propagating plasmons. Moreover, we demonstrate that by connecting the input and the output of the system, e.g., with a ring-shaped graphene-SiC nanostructure, it may be possible to spontaneously generate graphene SPPs (spasing [55-65]). It should be mentioned that the SPP amplification by means of a drift current biasing was studied in [66-67] in a related system, but with the effect of the drift current bias on the SPP waves treated simply adapting classical formulae from microwave theory to the graphene.

Figure 1 presents a schematic illustration of the structure under study. It consists of a graphene sheet biased with a drift electric current deposited on the top of a SiC substrate. We assume that the region above graphene is air. The frequency dispersion and dissipation in SiC are modeled by the dielectric function reported in [69-70]. In the absence of drifting electrons, the graphene sheet response may be characterized by the low-temperature nonlocal random-phase approximation (RPA) surface conductivity $\sigma_g(\omega,q)$ ($q=\sqrt{k_x^2+k_y^2}$ is the in-plane wavenumber) reported in Ref. [21], which includes both the intraband and interband contributions. For $\omega$ and $q$ complex, we evaluate $\sigma_g(\omega,q)$ using the analytical continuation formulas reported in Ref. [71]. The loss due to electronic scattering is modeled using the relaxation-time approximation [72]. We assume throughout this Letter the low-temperature limit (i.e., $k_B T \ll \mu_c$) and that the relaxation time in graphene is $\tau = 170$ fs [73-74], which is a conservative value compared to more recent observations [32]. Moreover, in the main text the chemical potential of the graphene sheet is taken equal to $\mu_c = 0.35$ eV. The space-time variation is assumed to be of the form $e^{ik_x x} e^{-i\omega t}$.

The graphene conductivity in the presence of a drift-current bias may be obtained from the conductivity without drift using a Galilean-Doppler shift [68, 75]

$$\sigma_g^{\text{drift}}(\omega, k_x) \approx (\omega/\tilde{\omega}) \sigma_g(\tilde{\omega}, q)\Big|_{q=\sqrt{k_x^2}}, \tag{1}$$

where $\tilde{\omega} = \omega - k_x v_0$ is the Doppler-shifted frequency and $k_x$ is the wave number along the $x$-direction. Here, $\sigma_g(\omega,q)$ is the nonlocal no-drift graphene conductivity discussed in the previous paragraph. It is assumed that the drifting electrons flow along the $x$-direction with drift velocity $v_0$ [see Fig. 1], and that the in-plane electric field is oriented along $x$ (longitudinal excitation). Remarkably, for sufficiently large positive $v_0$ and $k_x$, $\omega$ and $\tilde{\omega}$ have different signs, and consequently $\text{Re}\{\sigma_g^{\text{drift}}(\omega,k_x)\}$ may become negative in the upper-half frequency plane. Thereby, the drifting electrons may turn the graphene sheet into an active medium with optical gain [68] [see Fig. 2(a)].

Before discussing the loss compensation and plasmon amplification in the graphene-SiC waveguide, it is instructive to first examine the scattering properties of the drift-current biased graphene sheet when it is deposited on the top of a dielectric slab. To this end, we consider that a transverse magnetic (TM) wave with magnetic field directed along $y$ [see the inset of Fig. 2(b)] and characterized by the wave number $k_x$ illuminates the graphene sheet. The complex amplitude of the incident magnetic field is denoted by $H_y^{\text{inc}}$ and the (real-valued) oscillation frequency by $\omega$.

Due to the intrinsic material absorption, the superposition of the incident and reflected evanescent waves typically gives rise to a power flux towards the graphene sheet. The $z$-component of the total Poynting vector in the air region is given by $S_z = \frac{H_y^{\text{inc}}}{2\omega\varepsilon_0} \text{Im}\{\gamma_0(1-R)(1+R)^*\}$. Here,

$$R(\omega, k_x) = \frac{\gamma_0 \gamma_d + \kappa_g^{\text{drift}}(\gamma_d - \gamma_0 \varepsilon_{r,d})}{\gamma_0 \gamma_d - \kappa_g^{\text{drift}}(\gamma_d + \gamma_0 \varepsilon_{r,d})} \quad (2)$$

is the magnetic field reflection coefficient [76-77], $\kappa_g^{\text{drift}} = i\omega\varepsilon_0 / \sigma_g^{\text{drift}}$, $\varepsilon_0$ is the free-space permittivity, $\gamma_0 = \sqrt{k_x^2 - (\omega/c)^2}$ and $\gamma_d = \sqrt{k_x^2 - \varepsilon_{r,d}(\omega/c)^2}$ are the attenuation constants (along $z$) in the air and dielectric regions, respectively, $c$ is the speed of light in vacuum, and the "*" symbol denotes complex conjugation. Evidently, in the absence of a drift-current biasing the $z$-component of the Poynting vector is negative ($S_z < 0$) and thereby the graphene sheet absorbs energy (see the black curve in Fig. 2(b)). In contrast, with the drift-current biasing and for a large $k_x$ (blue and green curves in Fig. 2(b)), the energy density flux $S_z$ may flip its sign so that the graphene sheet may generate energy that flows away from it. This gain regime stems from the negative

Landau damping effect reported in [68], which enables the transfer of kinetic energy from the drifting electrons to the radiation field.

To study the opportunities created by the negative Landau damping effect, next we characterize the SPPs supported by the graphene-SiC system illustrated in Fig. 1. The dispersion characteristic of the SPPs is given by [70]

$$\frac{1}{\gamma_0} + \frac{\varepsilon_{r,SiC}(\omega)}{\gamma_{SiC}} - \frac{\sigma_g^{drift}}{i\omega\varepsilon_0} = 0, \quad (3)$$

where $\varepsilon_{r,SiC}(\omega)$ is the SiC dielectric function [69-70] and $\gamma_{SiC} = \sqrt{k_x^2 - \varepsilon_{r,SiC}(\omega/c)^2}$ is the attenuation constant (along $z$) of the plasmons in the SiC slab. If the drift-velocity is set identical to zero ($\sigma_g^{drift} \to \sigma_g$), one recovers the well-known dispersion equation for the plasmons supported by the graphene-SiC system.

Figure 3 depicts the dispersion characteristic of the SPPs supported by the structure for different drift velocities $v_0$. The dispersion is found by solving Eq. (3) with respect to $k_x = k_x' + ik_x''$ for real-valued $\omega$. For low-frequencies SiC behaves as a dielectric with positive permittivity. This happens for $\omega$ below the SiC resonance frequency $\omega_{TO}/(2\pi) = 22.78$ THz ($\omega_{TO}$ is the bulk transverse optical (TO) phonon frequency) [69-70]. In such a regime, the system is analogous to a graphene sheet placed on the top of a dielectric substrate, similar to the systems analyzed by us in [71, 76]. As shown in previous works [71, 76, 78-82], the drift-current biasing causes a symmetry breaking in the SPPs dispersion such that $\omega(k_x') \neq \omega(-k_x')$ [see inset (*i*) of Fig. 3(a)]. Similar nonreciprocal effects may also occur in systems with moving components [83-85]. Clearly, the degree of asymmetry increases with the drift velocity $v_0$ and, for sufficiently large $v_0$, it gives rise to regimes of unidirectional propagation wherein the SPPs are allowed to propagate only along the $+x$ direction (the direction of the drifting electrons) [71, 76].

On the other hand, for frequencies above the resonance and below $27.82\text{ THz}$, the real part of the SiC permittivity is negative ($\text{Re}\{\varepsilon_{\text{SiC}}\} < 0$) and thereby the SiC has a plasmonic (metal-type) response [69-70]. In the remainder of this Letter we focus our attention on that spectral range ($[22.78\text{ THz} - 27.82\text{ THz}]$).

The metal-phase of SiC, when $\text{Re}\{\varepsilon_{\text{SiC}}\} < 0$, leads to a pronounced spectral asymmetry of the graphene plasmons dispersion such that $\omega(k'_x) \neq \omega(-k'_x)$ [see Fig. 3(a), especially the insets (*ii*) and (*iii*)]. Even more interesting, Figs. 3(b)(*i*)-(*ii*) show that with the drift-current bias, the attenuation constant $\alpha = k''_x \text{sgn}(k'_x)$ of the SPPs co-propagating (counter-propagating) with the drifting electrons is greatly reduced (enhanced). Crucially, for large enough drift velocities $v_0$, the attenuation constant of the SPPs co-propagating with the drifting electrons ($k'_x > 0$) vanishes or even becomes negative. Specifically, Fig. 3(b)(*i*) shows that the graphene plasmons attenuation can be fully suppressed (i.e., $\alpha = 0$) for drift velocities on the order of $v_0 = v_F/4$ [see blue solid curve] and, for $v_0 > v_F/4$, it can be even overcompensated (i.e., $\alpha < 0$) [see green and purple solid curves]. Therefore, these results indicate that for $v_0 \geq v_F/4$, the graphene-SiC system may be either immune to attenuation or behave as an optical amplifier for the graphene plasmons copropagating with the drifting electrons. In contrast, for counter-propagating plasmons ($k'_x < 0$), $\alpha$ increases with the drift velocity $v_0$ [see dashed curves in Fig. 3(b)(*i*)], and hence, the drift current strongly suppresses the counter-propagating plasmons.

The SPP amplification strength and bandwidth increase with the drift velocity $v_0$. Curiously, the amplification strength $-\alpha$ for $v_0 = v_F/2$ may be comparable or even larger than the attenuation factor $\alpha$ of the SPPs without the drift-current biasing [see purple solid and black curves in Fig. 3(b)(*i*)]. Furthermore, the attenuation strength along the $-x$ direction for $v_0 = v_F/2$ is about 4 times larger than without drift [see purple and black dashed curves in Fig. 3(b)(*i*)]. On the other hand, Fig. 3(b)(*ii*) shows that the bandwidth of the SPP amplification regime is about 0.6 THz for $v_0 = v_F/3$ [see green curve], increasing up to around 1.2 THz as the drift velocity approaches $v_0 = v_F/2$ [see purple curve]. Interestingly, it is shown in Supplemental Material [77] that by increasing the chemical potential $\mu_c$ of graphene, one can boost the amplification bandwidth and the amplification gain $-\alpha$, and thereby reduce the threshold velocity $v_0$ at which the attenuation is fully suppressed ($\alpha = 0$). This happens because a larger $\mu_c$ implies a larger $k'_x$ (i.e., the SPPs are more confined), allowing that $\tilde{\omega}$, and consequently $\text{Re}\{\sigma_g^{\text{drift}}(\omega, k_x)\}$, become negative for lower drift velocities.

To further highlight the consequences of the loss compensation and gain regimes in the graphene-SiC system, next we consider the scenario wherein a linearly-polarized emitter (a short vertical electric dipole) placed in the vicinity of the graphene sheet is used to excite the graphene plasmons. The radiated and scattered fields are obtained from a Sommerfeld-type integral (an inverse Fourier-Laplace transform in $k_x$) as described in the Supplemental Material [77]. Because of the active response of the system, when $v_0 > 0$ the integral in $k_x$ must be calculated along a line in the lower half $k_x$-plane parallel to the real-$k_x$ axis. The integration path must be below all poles (for details see [77]).

Figure 4 shows the time snapshots of the *x*-component of the electric field for a graphene sheet biased with different drift velocities $v_0$. As expected, without a drift-current biasing ($v_0 = 0$), the two identical counter-propagating SPPs excited by the near-field emitter are equally attenuated as they propagate along the graphene-SiC interface [see Figs. 4(a-b)(*i*)]. In such circumstances, the SPP field attenuation is simply determined by the graphene and SiC damping rates. Quite differently, when a drift-current biasing is applied ($v_0 \neq 0$), the plasmons co-propagating with the drifting electrons (the +*x* direction) are significantly less attenuated than the plasmons propagating in the opposite direction (the −*x* direction) [see Fig. 4(a)(*ii*)]. In particular, for $v_0 = v_F/4$ the system supports loss-free plasmons that propagate along the +*x* direction [see Fig. 4(a)(*iii*) and Fig. 4(b)(*ii*)], which is consistent with the attenuation suppression predicted in Figs. 3(b)(*i*)-(*ii*). Notably, for drift velocities $v_0 > v_F/4$ the plasmons co-propagating with the drifting electrons (the +*x* direction) are amplified, whereas the plasmons propagating along the opposite direction are strongly attenuated [see Fig. 4(a)(*iv*) and Fig. 4(b)(*iii*)], as expected from the results of Figs. 3(b)(*i*)-(*ii*). As previously discussed, the optical gain is due to the conversion of kinetic energy of the drifting electrons into plasmon oscillations [68]. In Supplemental Material [77], we show that by increasing the chemical potential $\mu_c$, it is possible to reach regimes of lossless propagation and plasmon amplification for drift velocities even lower than $v_F/4$.

So far, it was assumed that the graphene-SiC guide has infinite length in the longitudinal direction (*x*-direction). Let us now consider finite-length nanostructures. In particular, let us consider a "circular" graphene resonator formed by a ring-shaped graphene ribbon with a drift-current bias placed on the top of a SiC substrate, as sketched in Fig. 5(a). The modes supported by such a ring-shaped graphene resonator can be found enforcing periodic boundary conditions. Specifically, if the perimeter of the resonator is $L$ then the allowed wave numbers are $k_\varphi = 2\pi n/L$, i.e., the wave number is necessarily real-valued. Thus, the natural modes of a circular resonator can be found by looking for complex-valued solutions $\omega = \omega(k_\varphi)$ of Eq. (3) with $k_\varphi = 2\pi n/L$ real-valued. The time variation is $e^{-i\omega t} = e^{-i\omega' t} e^{\omega'' t}$ with $\omega = \omega' + i\omega''$ the complex resonance frequency of the relevant mode. The system will be unstable if $\omega'' > 0$. Here, we neglect the curvature of the circular resonator, so that $\omega = \omega(k_\varphi)$ may be determined from the modal dispersion [Eq. (3)] of the associated planar geometry. Furthermore, the effect of the finite lateral width of the graphene ribbon is disregarded in our analysis. Using these approximations, we depict in Figs. 5(b)-(c) the real and imaginary parts of the oscillation frequency ($\omega$) as a function of the normalized $k_\varphi$ for different drift velocities $v_0$. Remarkably, Fig. 5(c) shows that for $v_0 > v_F/4$ the nano-ring resonator supports oscillations that grow exponentially with time ($\omega'' > 0$). The wave instabilities stem from the feedback that is obtained by connecting the input and output of the optical amplifier. As expected, the growth rate increases as the drift velocity $v_0$ increases, and for $v_0 = v_F/2$, the growth rate (i.e., the magnitude of $\omega''$) can be as high as $\omega'' \approx 1.13 \times 10^{12}$ s$^{-1}$. Therefore, the graphene nanoresonator can be used as spaser pumped by drifting electrons [55-65].

In Supplemental Material [77], it is shown that the relativistic Doppler-shift model for the graphene conductivity [78-80, 86-87] leads to qualitatively similar wave instabilities but with a slightly larger growth rate. In our understanding the Galilean-Doppler shift theory is the one that models more accurately the drift-current bias when the electron-electron interactions predominate and force the electrons to move with constant velocity $v_0$ [75]. In addition, it is also shown in Supplemental Material [77] that the wave instabilities are rooted in the intraband light-matter interactions.

In summary, we have demonstrated that a system formed by a drift-current biased graphene sheet deposited on a SiC substrate may enable the full compensation of plasmonic losses and the amplification of graphene plasmons. The plasmonic gain is due to the conversion of the kinetic energy of the moving electrons into short-wavelength plasmons. It was shown that a graphene-SiC waveguide behaves as a one-way optical amplifier. Finally, it was demonstrated that a ring-shaped graphene-SiC resonator can be used as spaser pumped by the drifting electrons.

**Acknowledgments:** This work was partially funded by the IET under the A F Harvey Prize, by the Simons Foundation under the award 733700 (Simons Collaboration in Mathematics and Physics, "Harnessing Universal Symmetry Concepts for Extreme Wave Phenomena"), and by Instituto de Telecomunicações (IT) under project UIDB/50008/2020. T. A. Morgado acknowledges FCT for research financial support with reference CEECIND/04530/2017 under the CEEC Individual 2017, and IT-Coimbra for the contract as an assistant researcher with reference CT/No. 004/2019-F00069.

**Figures**

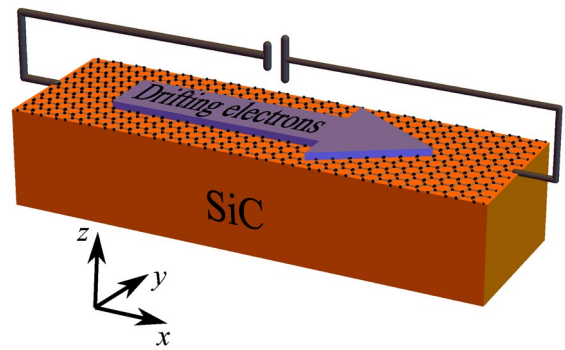

**Fig. 1.** A graphene sheet deposited on the top of a SiC substrate is biased with a drift-electric current due to a static voltage generator.

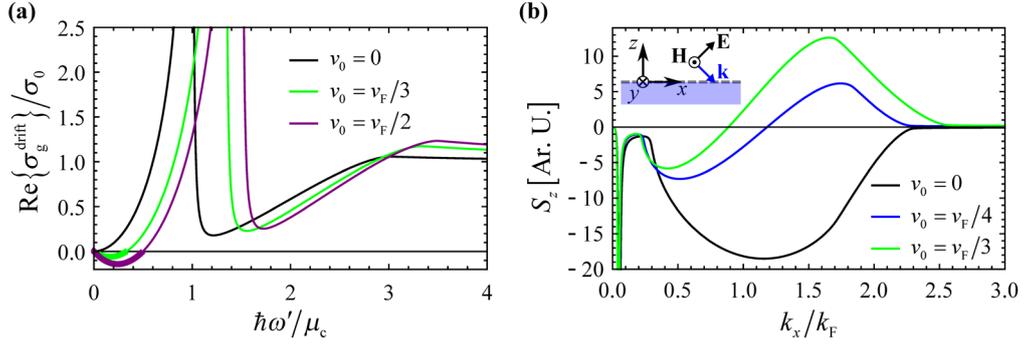

**Fig. 2.** (a) Real part of the graphene conductivity in the upper-half frequency-plane as a function of the normalized frequency $\hbar\omega'/\mu_c$ ($\omega = \omega' + i\omega''$) for different drift velocities, $k_x = k_F$ and $\omega''/(2\pi) = 0.1\,\text{THz}$. The thicker green and purple curves correspond to the gain regions. The conductivity normalization factor is $\sigma_0 = e^2/(4\hbar)$. (b) Poynting vector component perpendicular to the interface (in arbitrary units) as a function of $k_x$ for different drift velocities and $f = 25\,\text{THz}$. The inset depicts a TM plane wave illuminating the drift-current biased graphene sheet deposited on a dielectric with relative permittivity $\varepsilon_{r,d} = 4$.

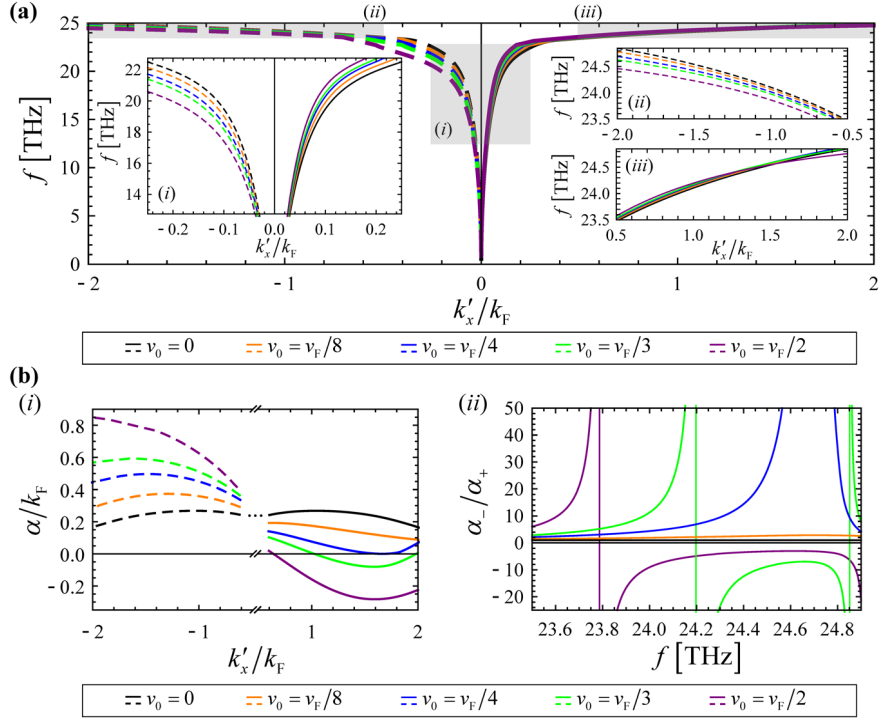

**Fig. 3.** Dispersion of the SPPs supported by the graphene-SiC system for several drift velocities $v_0$. (a) Main panel: SPP frequency as a function of the real part of the SPP wavenumber; (*i*), (*ii*), and (*iii*) zoom-in views of the shaded rectangular areas of the main panel. (b) (*i*) SPP attenuation constant ($\alpha = k_x'' \operatorname{sgn}(k_x')$) as a function of the real part of the SPP wavenumber. (*ii*) Ratio between the attenuation constants of the SPPs that propagate along the $-x$ and $+x$ direction as a function of the frequency.

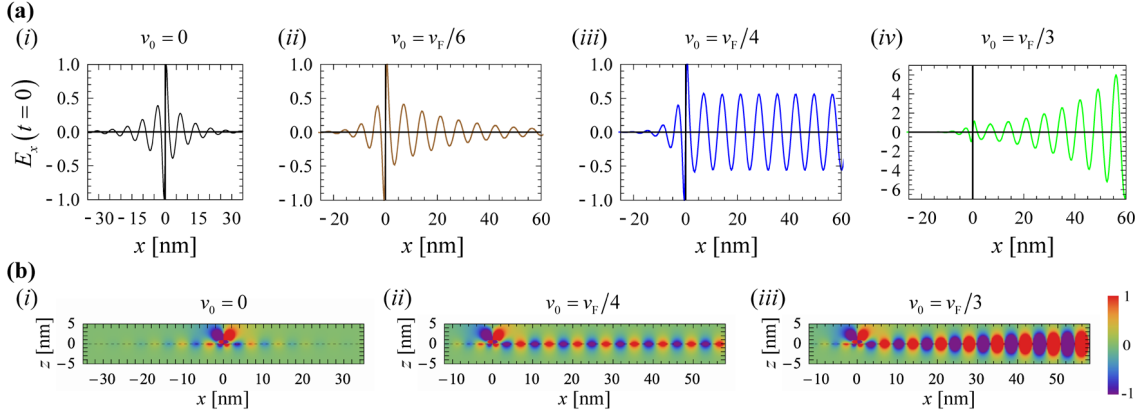

**Fig. 4.** SPP excitation by a near-field emitter. (a) Time snapshots of the *x*-component of the electric field $E_x$ (in arbitrary unities) as a function of $x$ and for $z = 0$, for several drift velocities $v_0$. (b) Time snapshots of the *x*-component of the electric field $E_x$ as a function of $x$ and $z$ and for several drift velocities $v_0$. The frequency of operation is $\omega/(2\pi) = 24.7$ THz and the emitter is positioned at the point $(x, z) = (0, 1 \text{ nm})$. The drift velocity $v_0$ is indicated at the top of each panel.

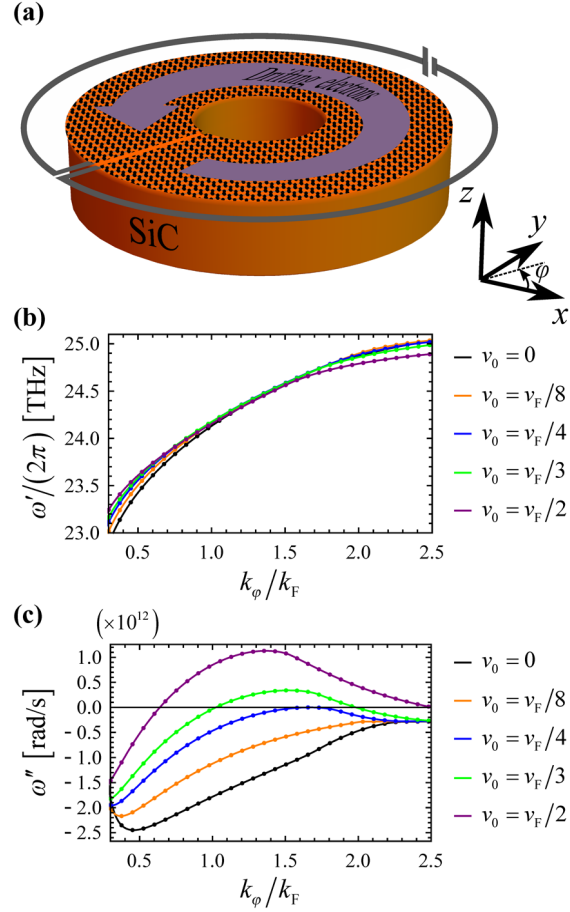

**Fig. 5.** (a) Circular graphene nanoresonator formed by a ring-shaped graphene ribbon biased with a drift current. (b) Real and (c) imaginary parts of the oscillation frequencies of the natural modes as a function of $k_\varphi$ for different drift velocities $v_0$. Discrete points: oscillation frequencies for a circular graphene nanoresonator with radius $R = 25$ nm; Solid lines: oscillation frequencies for a circular resonator with $R \to \infty$. The spasing occurs for modes with $\omega'' > 0$.

# Supplemental Material for
# "Active graphene plasmonics with a drift-current bias"

Tiago A. Morgado[1], Mário G. Silveirinha[1,2*]

[1]*Instituto de Telecomunicações and Department of Electrical Engineering, University of Coimbra, 3030-290 Coimbra, Portugal*

[2]*University of Lisbon, Instituto Superior Técnico, Avenida Rovisco Pais, 1, 1049-001 Lisboa, Portugal*

*E-mail:* tiago.morgado@co.it.pt, mario.silveirinha@co.it.pt

In the supplementary note *A*) we study the impact of the chemical potential $\mu_c$ in the dispersion and field enhancement of the SPPs supported by the graphene-SiC system with a drift-current bias. In the supplementary note *B*), we find the reflection and transmission coefficients for the graphene-SiC interface and a plane wave excitation. In the supplementary note *C*) we derive the electromagnetic fields radiated by a linearly-polarized emitter placed above the graphene-SiC system. Finally, in the supplementary note *D*) we compare the amplification strength obtained with our conductivity model with what is predicted by the "relativistic Doppler-shift" conductivity model.

## *A. Impact of the chemical potential in the SPP dispersion and field enhancement*

Figure S1 depicts the dispersion of the SPPs supported by the graphene-SiC system (analogous to Fig. 3 of the main text) for the chemical potentials (a) $\mu_c = 0.2$ eV and (b) $\mu_c = 0.5$ eV. As seen from Fig. S1(a)(*ii*) and Fig. S1(b)(*ii*), the magnitude of the amplification (attenuation) coefficient $-\alpha$ ($\alpha$) of the SPPs copropagating (counterpropagating) with the drifting electrons is significantly enhanced by increasing the chemical potential $\mu_c$. It can be seen that for the same drift velocity $v_0$, $-\alpha$ ($\alpha$) of the SPPs copropagating (counterpropagating) with the streaming current carriers is considerably larger in the system with higher $\mu_c$ [see, for instance, the purple solid and dashed curves in Fig. S1(a)(*ii*) and Fig. S1(b)(*ii*)].

---

[*] To whom correspondence should be addressed: E-mail: mario.silveirinha@co.it.pt

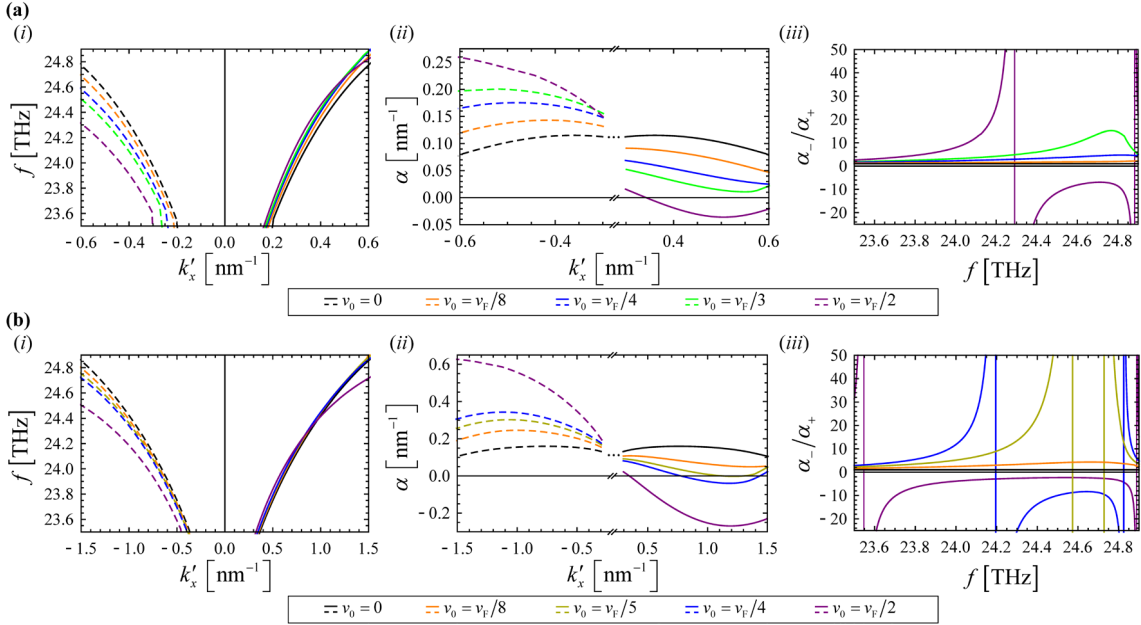

**Fig. S1.** Dispersion of the SPPs supported by the graphene-SiC system for the chemical potentials. (a) $\mu_c = 0.2$ eV; (b) $\mu_c = 0.5$ eV. (*i*) SPP frequency as a function of the real part of the SPP wavenumber; (b) (*i*) SPP attenuation constant ($\alpha = k_x'' \operatorname{sgn}(k_x')$) as a function of the real part of the SPP wavenumber. (*iii*) Ratio between the attenuation constants of the SPPs that propagate along the $-x$ and $+x$ direction as a function of the frequency.

Figures S1(a)(*iii*)-(b)(*iii*) show that the bandwidth of the SPP amplification regime (range of frequencies where $\alpha_+ < 0$) also increases with $\mu_c$. For $v_0 = v_F/2$ the amplification bandwidth is about 0.6 THz for $\mu_c = 0.2$ eV [see the purple solid in Fig. S1(a)(*iii*)], increasing up to 1.35 THz for $\mu_c = 0.5$ eV [see the purple solid in Fig. S1(b)(*iii*)].

Figure S2 confirms that graphene-SiC systems with higher $\mu_c$ enable SPP amplification for smaller drift velocities. For example, for $\mu_c = 0.2$ eV the SPP amplification regime is only attainable for $v_0 > v_F/3$, whereas for $\mu_c = 0.5$ eV the amplification threshold velocity can be as small as $v_0 = v_F/5$.

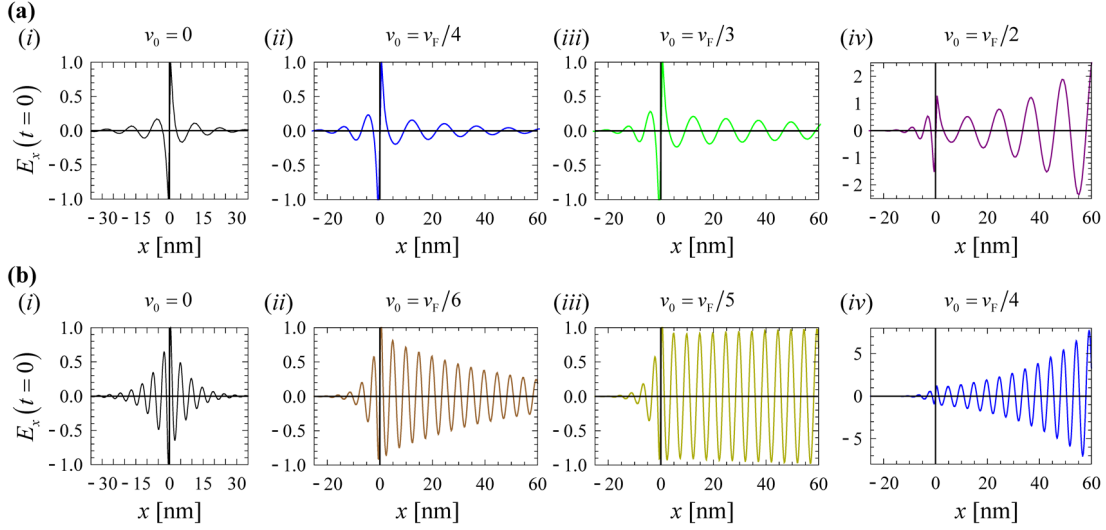

**Fig. S2.** (a) Time snapshots of the x-component of the electric field $E_x$ (in arbitrary unities) as a function of $x$ and for $z = 0$, for several drift velocities $v_0$ and two different chemical potentials. (a) $\mu_c = 0.2$ eV; (b) $\mu_c = 0.5$ eV. The frequency of operation is $f = 24.7$ THz. The value of the drift velocity $v_0$ is indicated explicitly in the top of each panel.

## B. Reflection and transmission coefficients

The reflection and transmission coefficients for a transverse magnetic (TM) polarized wave incident on a drift-current biased graphene sheet (see Fig. S3) can be obtained in the usual way by expanding the electromagnetic field in all the regions of space in terms of plane waves, and then solving for the unknown wave amplitudes with mode matching.

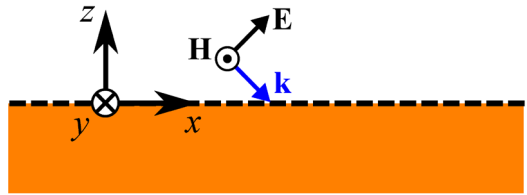

**Fig. S3.** Sketch of a graphene sheet (with a drift-current bias) deposited above the SiC substrate and illuminated by a TM-polarized wave.

Assuming that the incident magnetic field is along the y-direction and has complex amplitude $H_y^{\text{inc}}$, it follows that the magnetic field in all space can be written as (the variation $e^{i(k_x x - \omega t)}$ of the fields is omitted):

$$H_y = H_y^{inc}\left(e^{\gamma_0 z} + Re^{-\gamma_0 z}\right), \quad z > 0$$
$$H_y = H_y^{inc} T e^{\gamma_{SiC} z}, \quad z < 0$$

(S1)

In the above, $\gamma_0 = \sqrt{k_x^2 - (\omega/c)^2}$ and $\gamma_{SiC} = \sqrt{k_x^2 - \varepsilon_{r,SiC}(\omega/c)^2}$ are the propagation constants along $z$ in the vacuum and SiC regions, and $R$ and $T$ are the reflection and transmission coefficients, respectively. The electric field distribution can be easily found using the Maxwell's equations. By matching the tangential component of the electric field ($E_x|_{z=0^+} - E_x|_{z=0^-} = 0$) and by imposing the impedance boundary condition ($H_y|_{z=0^+} - H_y|_{z=0^-} = -\sigma_g^{drift} E_x$) at the interface [1-2], it is found that the reflection and transmission coefficients satisfy

$$R(\omega, k_x) = \frac{\gamma_0 \gamma_{SiC} + \kappa_g^{drift}\left(\gamma_{SiC} - \gamma_0 \varepsilon_{r,SiC}\right)}{\gamma_0 \gamma_{SiC} - \kappa_g^{drift}\left(\gamma_{SiC} + \gamma_0 \varepsilon_{r,SiC}\right)}$$

$$T(\omega, k_x) = \frac{-2\kappa_g^{drift} \gamma_0 \varepsilon_{r,SiC}}{\gamma_0 \gamma_{SiC} - \kappa_g^{drift}\left(\gamma_{SiC} + \gamma_0 \varepsilon_{r,SiC}\right)}$$

(S2)

where $\sigma_g^{drift}(\omega, k_x) = (\omega/\tilde{\omega})\sigma_g(\tilde{\omega})$ is the graphene conductivity in the presence of a drift-current bias, $\tilde{\omega} = \omega - k_x v_0$ is the Doppler-shifted frequency, $\sigma_g(\omega)$ is the nonlocal graphene conductivity [3-4] and $\kappa_g^{drift} = i\omega\varepsilon_0 / \sigma_g^{drift}$. Evidently, when $v_0 = 0$ the reflection and transmission coefficients reduce to the standard formulas in the absence of a drift-current bias [1].

## C. Fields radiated by a linearly polarized emitter above the graphene-SiC system

We consider a short emitter (polarized along the vertical direction) characterized by the current density $\mathbf{j}_e = -i\omega p_e \delta(x - x_0)\delta(z - z_0)\hat{\mathbf{z}}$, with $p_e$ the electric dipole moment per unit of length. The emitter is placed in vacuum at a distance $z_0$ from the graphene-SiC interface [see Fig. S4]. For simplicity, we consider that the current density is independent of the $y$-coordinate so that the problem is two-dimensional (2D). Looking for a solution of the Maxwell's equations

$$\nabla \times \mathbf{E} = i\omega\mu_0 \mathbf{H}, \qquad \nabla \times \mathbf{H} = -i\omega\varepsilon_0 \mathbf{E} + \mathbf{j}_e, \qquad (S3)$$

of the form $\mathbf{H} = H_y(x,z)\hat{\mathbf{y}}$ it is found that for $z > 0$:

$$\nabla^2 H_y + \left(\frac{\omega}{c}\right)^2 H_y = -i\omega p_e \frac{\partial}{\partial x}\left[\delta(x-x_0)\delta(z-z_0)\right]. \qquad (S4)$$

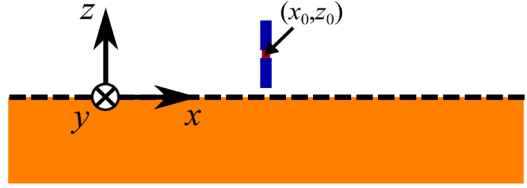

**Fig. S4.** A linearly polarized emitter is placed at the position $(x,z) = (x_0, z_0)$ above the graphene-SiC interface.

In a scenario wherein the emitter is embedded in a bulk (unbounded) medium (without the graphene sheet) the radiated field is given by:

$$H_y^{\text{inc}} = i\omega p_e \frac{\partial}{\partial x}\left[-\frac{1}{4i} H_0^{(1)}\left(\frac{\omega}{c}\sqrt{\varepsilon_{r,h}}\, r\right)\right]$$
$$= -\omega p_e \frac{1}{2\pi}\int \frac{k_x}{2\gamma_h} e^{-\gamma_h|z-z_0|} e^{ik_x(x-x_0)} dk_x, \qquad (S5)$$

where $H_0^{(1)}$ is the Hankel function of first kind and order zero, $r = \sqrt{(x-x_0)^2 + (z-z_0)^2}$, $\varepsilon_{r,h}$ is the relative permittivity of the host medium, $\gamma_h = -i\sqrt{(\omega/c)^2 \varepsilon_{r,h} - k_x^2}$, and the integration range is over the entire real axis. The integral representation of the emitted field can be straightforwardly adapted to take into account for the presence of the graphene sheet at $z = 0$:

$$H_y = -\omega p_e \frac{1}{2\pi} \begin{cases} \int \frac{k_x}{2\gamma_0} \left( e^{-\gamma_0 |z-z_0|} + R e^{-\gamma_0(z+z_0)} \right) e^{ik_x(x-x_0)} dk_x, & z > 0 \\ \int \frac{k_x}{2\gamma_{\text{SiC}}} T e^{\gamma_{\text{SiC}}(z-z_0)} e^{ik_x(x-x_0)} dk_x, & z < 0 \end{cases}, \quad (S6)$$

where $\gamma_0 = -i\sqrt{(\omega/c)^2 - k_x^2}$ and $\gamma_{\text{SiC}} = -i\sqrt{(\omega/c)^2 \varepsilon_{\text{r,SiC}} - k_x^2}$ are the propagation constants along $z$ in the vacuum and SiC regions, respectively, $\varepsilon_{\text{r,SiC}}(\omega)$ is the dielectric function of the SiC [1, 5], and $R = R(\omega, k_x)$ and $T = T(\omega, k_x)$ represent the (magnetic field) reflection and transmission coefficients for transverse magnetic (TM)-polarized waves, which are given in Sect. B. The $x$- and $z$- components of the electric field can be obtained from the Maxwell-Ampère equation so that:

$$E_x = \frac{ip_e}{\varepsilon_0} \frac{1}{2\pi} \begin{cases} \int \frac{k_x}{2} \left( -\text{sgn}(z-z_0) e^{-\gamma_0|z-z_0|} - R e^{-\gamma_0(z+z_0)} \right) e^{ik_x(x-x_0)} dk_x, & z > 0 \\ \frac{1}{\varepsilon_{\text{r,SiC}}} \int \frac{k_x}{2} T e^{\gamma_{\text{SiC}}(z-z_0)} e^{ik_x(x-x_0)} dk_x, & z < 0 \end{cases}, \quad (S7)$$

$$E_z = \frac{p_e}{\varepsilon_0} \frac{1}{2\pi} \begin{cases} \int \frac{k_x^2}{2\gamma_0} \left( e^{-\gamma_0|z-z_0|} + R e^{-\gamma_0(z+z_0)} \right) e^{ik_x(x-x_0)} dk_x, & z > 0 \\ \frac{1}{\varepsilon_{\text{r,SiC}}} \int \frac{k_x^2}{2\gamma_{\text{SiC}}} T e^{\gamma_{\text{SiC}}(z-z_0)} e^{ik_x(x-x_0)} dk_x, & z < 0 \end{cases}. \quad (S8)$$

In the amplification regime (which corresponds to a convective instability), the integration path in Eqs. (S6-S8) must be displaced from the real axis to the $k_x$-complex plane by adding to $k_x$ an imaginary component $i\delta$ (i.e., $k_x \to k_x + i\delta$). For $v_0 > 0$ the fields are expected to grow exponentially when $x \to +\infty$. In this case, the Fourier (Laplace) transform of the fields with respect to $x$ is well-defined for $\text{Im}\{k_x\} < 0$. Therefore, the inverse Fourier transform must be done through a contour with $\text{Im}\{k_x\} < 0$ below all poles. Hence, for $v_0 > 0$, the value of $\delta$ must be negative, with a magnitude larger than the peak value of the amplification gain $(-\alpha)_{\max}$ (for $v_0 < 0$ the value of $\delta$ is positive).

## D. Comparison between different graphene conductivity models

In the main text, the drift-biased graphene conductivity was evaluated using the Galilean Doppler shift model introduced by us in [6]. Alternative models have been proposed by other research groups [7-12]. The physics predicted by such models is well captured by a *relativistic* Doppler transformation such that

$$\sigma_{\text{g}}^{\text{Rel}}(\omega, k_x) \approx (\omega/\tilde{\omega})\sigma_{\text{g}}(\tilde{\omega}, \tilde{q})\big|_{\tilde{q}=\sqrt{\tilde{k}_x^2}}, \qquad (S9)$$

with $\tilde{\omega} = \gamma(\omega - k_x v_0)$, $\tilde{k}_x = \gamma(k_x - \omega v_0/v_{\text{F}}^2)$, and $\gamma = 1/\sqrt{1 - v_0^2/v_{\text{F}}^2}$ the graphene Lorentz factor. For further details the reader is referred to our previous works [4, 13].

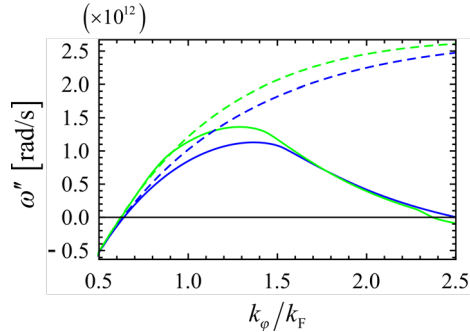

**Fig. S5.** Imaginary part of the oscillation frequencies of the natural modes of a circular graphene nanoresonator with $R \to \infty$ and $v_0 = v_{\text{F}}/2$, analogous to Fig. 5 of the main text. The results are calculated with our Galilean Doppler shift model (blue curves) and the relativistic Doppler shift model [Eq. (S9)] (green curves). The solid lines were obtained with the (bare) conductivity formula that accounts for both intraband and interband contributions (Eq. (A2) of Ref. [4]), whereas the dashed lines were calculated using the (bare) conductivity formula that includes only the intraband contribution (Eq. (C2) of Ref. [4]).

Figure S5 depicts the growth rate of the unstable mode of a circular graphene nanoresonator with $R \to \infty$ as a function of the azimuthal wave number $k_\varphi$ (analogous to Fig. 5 of the main text). The results were obtained using our Galilean Doppler shift model (blue solid curves) and the relativistic Doppler shift model [Eq. (S9)] (green solid curves) with a (bare) graphene conductivity that takes into account both intraband and interband contributions. It is seen that the relativistic Doppler shift model predicts wave instabilities with a slightly larger magnitude than our Galilean Doppler shift model.

In addition, we show in Fig. S5 [dashed curves] the growth rate calculated using the Galilean and relativistic Doppler-shift models with the (bare) intraband graphene conductivity. The wave instabilities predicted by the intraband models have strength comparable to those obtained using the full (intraband + interband) graphene response for $k_\varphi < k_F$. For larger values of $k_\varphi$, the intraband models overestimate the instability strength. This is due to the optical loss associated with the interband transitions. Thus, generically speaking, the interband light-matter interactions have a minor influence on the emergence of unstable regimes.